\documentclass[reprint, amsmath, amssymb, aps, prl]{revtex4-2}
\usepackage{amsmath}
\usepackage{graphicx}

\makeatletter
\DeclareMathOperator{\sech}{sech}
\usepackage{dcolumn}
\usepackage{bm}
\usepackage{xcolor}

\makeatother

\begin{document}
\title{Creation of two-mode squeezed states in atomic mechanical oscillators}
\author{Wui Seng Leong$^1$}
\author{Mingjie Xin$^1$}
\email{mjxin@ntu.edu.sg}

\author{Zilong Chen$^1$, Yu Wang$^1$}
\author{Shau-Yu Lan$^{1,2,3}$}%
\email{sylan@ntu.edu.tw}

\affiliation{%
$^{1}$Division of Physics and Applied Physics, School of Physical and Mathematical Sciences, Nanyang Technological University, Singapore 637371, Singapore\\
$^{2}$Department of Physics, National Taiwan University, Taipei 10617, Taiwan\\
$^{3}$Center for Quantum Science and Engineering,National Taiwan University, Taipei 10617, Taiwan}

\date{\today}

\begin{abstract}
Two-mode squeezed states, which are entangled states with bipartite quantum correlations in continuous-variable systems, are crucial in quantum information processing and metrology. Recently, continuous‐variable quantum computing with the vibrational modes of trapped atoms has emerged with significant progress, featuring a high degree of control in hybridizing with spin qubits. Creating two-mode squeezed states in such a platform could enable applications that are only viable with photons. Here, we experimentally demonstrate two‐mode squeezed states by employing atoms in a two‐dimensional optical lattice as quantum registers. The states are generated by a controlled projection conditioned on the relative phase of two independent squeezed states. The individual squeezing is created by sudden jumps of the oscillators’ frequencies, allowing generating of the two‐mode squeezed states at a rate within a fraction of the oscillation frequency. We validate the states by entanglement steering criteria and Fock state analysis. Our results can be applied in other mechanical oscillators for quantum sensing and continuous‐variable quantum information.
\end{abstract}
\pacs{Valid PACS appear here}

\maketitle
The Heisenberg uncertainty principle allows one to \textquotedblleft squeeze\textquotedblright{}
the noise of one quadrature below the vacuum fluctuation at the expense
of increasing the noise of its non-commuting quadrature, creating
quantum correlations between the quadratures. The squeezing, when
created between non-commuting quadratures of two modes, can be used
to create cross-correlations between them while the modes contain
no trace of correlations in their own quadratures. Such a two-mode
entangled state is also known as an Einstein\nobreakdash-Podolsky-Rosen
(EPR) state. In photonic systems, two-mode squeezed states have been
playing a pivotal role in optical quantum communications, such as
teleportation, dense coding, and quantum repeaters \citep{Bra,Wee,Fuk}.
Moreover, it has also been used to generate non-Gaussian states and
prepare cluster states for universal quantum computation \citep{Bra,Wee,Fuk}.
The implementation of continuous-variable quantum information with
vibrational modes of atoms, such as trapped ions, has been considered
favourably because of its flexibility and controllability in hybridising
with atomic spin states \citep{Che,Sut,Ort,And}. It has been used
to prepare Gottesman-Kitaev-Preskill (GKP) state \citep{Flu}, perform
error corrections \citep{de}, and demonstrate hybrid and non-Gaussian
operations \citep{Um,Gan}.

Proposals to generate a two-mode squeezed state in trapped atom systems
have used auxiliary oscillators or qubits as \textquotedblleft quantum
bus\textquotedblright{} mediators to perturbatively entangle two modes
of motion while creating squeezing at the same time \citep{Sut,Ort,Car,Bur}.
As a weak coupling is often necessary to avoid higher-order excitations,
those methods require significant time and resource overhead. On the
other hand, in the optical domain, overlapping two out-of-phase independent
squeezed states on a physical beam splitter is an efficient way to
create two-mode entanglement \citep{Fuk} without any direct interaction
of the two modes. In mechanical oscillators, one can simulate such
a physical beam splitter by projecting two spatially orthogonal oscillators
onto a 45-degree basis as the two output ports of the beam splitter.
Here, we utilise atoms trapped in an isotropic two-dimensional potential
to realise an interaction-free two-mode squeezed gate. Squeezing of
each mode is initialised independently along two orthogonal axes by
implementing sudden jumps of the oscillator frequencies \citep{Xin}.
The two-mode entanglement is verified by satisfying the EPR steering
criteria when the two input independent squeezed states are out of
phase. Conditioning on the in-phase of the two initial squeezed states,
a two-dimensional single-mode squeezed state can also be realized.

\begin{figure*}
\includegraphics[scale=0.55]{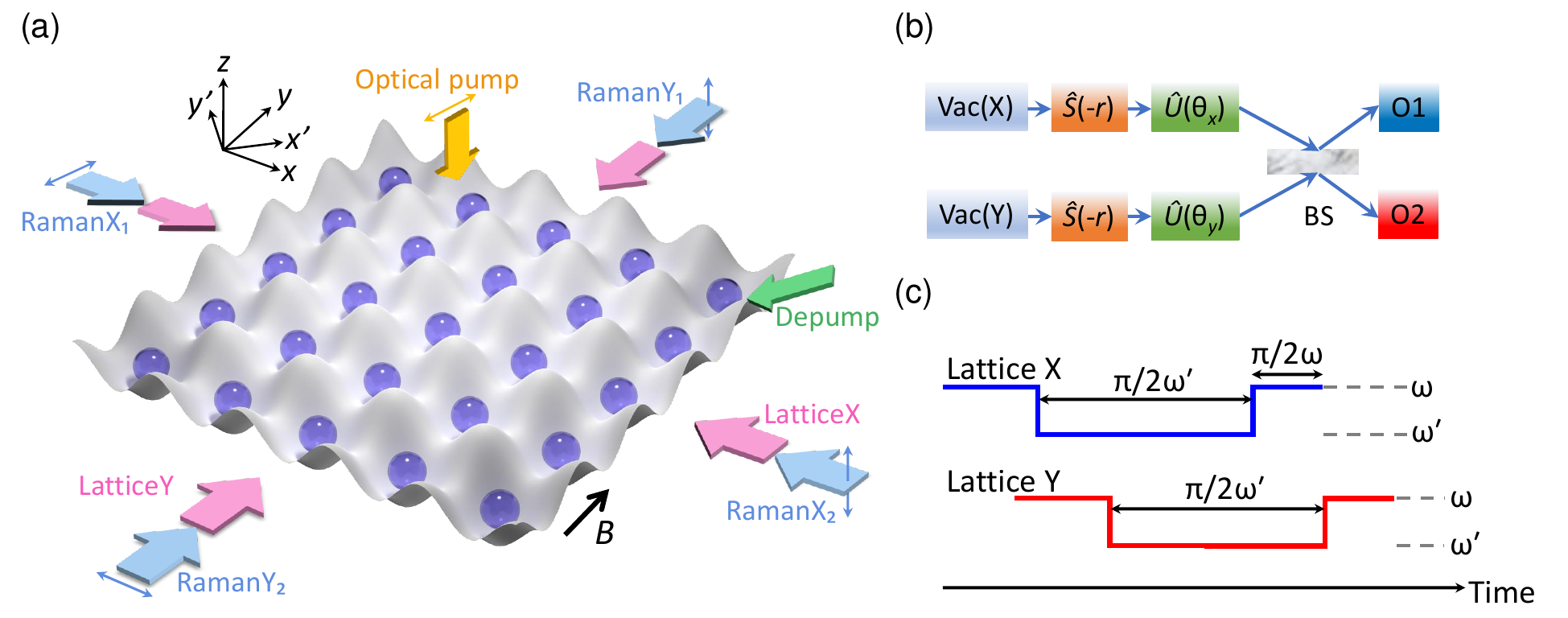}

\caption{\label{fig:Fig1}Illustration and experimental realization of two-mode
squeezing gates in single atoms oscillators. (a) Configuration of
the preparation and measurement of single atoms oscillators in two-dimensional
($x$--$y$ plane) quantum registers. Each ball represents a single
atom in a two-dimensional harmonic potential. The magnetic field $B$
(along the $y$ direction), optical pump and depump beams are used in resolved
Raman sideband cooling. Colored thin arrows indicate the polarization
of the optical beams. The Raman beams are co-propagating with the
lattice beams. (b) Schematic diagram of the two-mode squeezing quantum
logic. The vacuum (Vac) states in the $x$ and $y$ modes are independently
squeezed by a squeezing operator $\hat{S}\left(-r\right)$. The output
states O1 and O2 are conditioned on the free evolution operation $\hat{U}$
of the modes before the beam splitter (BS). When $\theta_{x}-\theta_{y}=\pi/2$, the output states are correlated as an EPR state. When $\theta_{x}-\theta_{y}=0$,
the output states are a two-dimensional single-mode squeezed state.
(c) Timing sequence of setting $\theta_{x}-\theta_{y}=\pi/2$ for
two-mode squeezed states. The independent squeezed states are generated
by jumping the oscillator frequency between $\omega$ and $\omega^{\prime}$.}
\end{figure*}

Consider two independent squeezing operators $\hat{S}_{x}(\xi_{x})$ and $\hat{S}_{y}(\xi_{y})$ along the $x$ and $y$
directions with squeezing parameters $\xi_{x}=re^{i2\theta_{x}}$
and $\xi_{y}=re^{i2\theta_{y}}$, the transformation of their annihilation
operators $\hat{a}_{x}$ and $\hat{a}_{y}$ under squeezing can be
expressed in terms of operators
\begin{equation}
\begin{split}\hat{A}_{x} & = \hat{S}^{\dag}_{x}(\xi_{x})\hat{a}_{x}\hat{S}_{x}(\xi_{x}) = \hat{a}_{x}\cosh r-e^{i2\theta_{x}}\hat{a}_{x}^{\dagger}\sinh r\\
\hat{A}_{y} & = \hat{S}^{\dag}_{y}(\xi_{y})\hat{a}_{y}\hat{S}_{y}(\xi_{y}) = \hat{a}_{y}\cosh r-e^{i2\theta_{y}}\hat{a}_{y}^{\dagger}\sinh r,
\end{split}
\end{equation}
where $r$ is the squeezing amplitude, $\theta_{x}=\omega t$ and
$\theta_{y}=\theta_{0}+\omega t$ are the phases of the two oscillators,
$\theta_{0}$ is the initial relative phase of two oscillators, $\omega$
is the isotropic oscillator frequency, $t$ is the free evolution
time, and $\hat{a}_{x}^{\dagger}$ and $\hat{a}_{y}^{\dagger}$ are
the creation operators. Upon a 50/50 beam splitter, the operators
of the two output ports $\hat{A}_{x}^{\prime}$ and $\hat{A}_{y}^{\prime}$
can be written as $\hat{A}_{x}^{\prime}=(\hat{A}_{x}+\hat{A}_{y})/\sqrt{2}$
and $\hat{A}_{y}^{\prime}=(-\hat{A}_{x}+\hat{A}_{y})/\sqrt{2}$. The
uncertainty of the momentum operators $\hat{p}_{x}^{\prime}=-i \Delta p_{0}(\hat{A}^{'}_{x}-\hat{A}^{'\dag}_{x})$ and
$\hat{p}_{y}^{\prime}=-i\Delta p_{0}(\hat{A}^{'}_{y}-\hat{A}^{'\dag}_{y})$ in this new basis can then be calculated as
\begin{equation}
\begin{split}\Delta p_{x(y)}^{\prime} & =\left\langle 0\left|\left(\hat{p}_{x(y)}^{\prime}\right)^{2}\right|0\right\rangle^{1/2}\\
 & =\Delta p_{0}\left[\cosh2r+\sinh2r\cos\left(\theta_{x}+\theta_{y}\right)\right.\\
 & \quad\times\left.\cos\left(\theta_{x}-\theta_{y}\right)\right]^{1/2},
\end{split}
\label{eq:En2}
\end{equation}
where $\Delta p_{0}$ is the uncertainty of the ground state momentum.
Conditioning on $\theta_{x}-\theta_{y}=\pi/2$ , the uncertainty of
the momentum at the outputs shows no correlation on the individual
modes, manifested by a constant width cosh2$r$ \citep{Eke}. When $\theta_{x}-\theta_{y}=0$, $\Delta p_{x(y)}^{\prime}=\Delta p_{0}\left[\cosh2r+\sinh2r\cos(2\omega t)\right]^{1/2}$
exhibit a variation of single-mode squeezed state at a rate of $2\omega$.

The two-dimensional harmonic potential in this experiment is formed
by two retro-reflected 1064 nm lasers aligned perpendicularly. Each
beam has a waist of 60 $\mu$m and 1 W of power, creating a two-dimensional
optical lattice with an isotropic peak trapping frequency $\omega$
= 2$\pi$ $\times$ 125 kHz as quantum registers, shown in Fig. \ref{fig:Fig1}(a).
An ensemble of cold $^{85}$Rb atoms is loaded into the optical lattice
after sub-Doppler cooling and then compressed by shuffling the two
lattice beams to minimize the inhomogeneous trapping frequency (see Supplemental Material). The ensemble of $6\times10^{4}$ atoms inside the trap exhibits a Gaussian distribution with a full width at half maximum (FWHM) along the $x$ and $y$ directions of 47 $\mu$m and 64 $\mu$m, respectively. After which, atoms are cooled down to the two-dimensional
ground states by resolved Raman sideband cooling \citep{Leo} (see Supplemental Material). During absorption detection, we selectively image a 30 $\mu$m
$\times$ 30 $\mu$m area of atoms along the $x$ direction to sample
a region with reduced inhomogeneous broadening of the oscillation
frequency.

Our protocol for two-mode squeezing generation is illustrated in Fig.
\ref{fig:Fig1}(b). The wave functions of the ground states in both
directions are squeezed independently by jumping the oscillator frequencies
\citep{Xin} between $\omega$ and $\omega^{\prime}$ through sudden
changes in the optical lattice power. The frequency jump determines
the squeezing amplitude $r$ as $\ln\left(\omega/\omega^{\prime}\right)$.
Such a method has been used to create nearly instantaneous operations
of squeezing \citep{Xin}. To prepare two squeezed states with a $\pi$/2
phase difference, the two independent squeezing operations are relatively
delayed by a time $\pi/\left(2\omega\right)$, as shown in Fig. \ref{fig:Fig1}(c).
Along the 45-degree basis, each of the new position observable $\hat{x}^{\prime}$
and $\hat{y}^{\prime}$ are superpositions of observables $\hat{x}$
and $\hat{y}$, which simulates a physical beam splitter. The output
modes along these two directions are conventionally named Alice and
Bob, who possess quantum objects that are correlated with each other,
while the quadratures of their own show no correlations. This is illustrated
by the Wigner function \citep{Eke} projected into the $p_{x}^{\prime}-x^{\prime}$
plane, as shown in Fig. \ref{fig:Fig2}(a).

We measure the uncertainty of velocity $v_{x}^{\prime}=p_{x}^{\prime}/m$
using two-photon Raman velocimetry with different magnitudes of frequency
jump, where $m$ is mass (see Supplemental Material). In this new basis, no correlation
between the quadratures is indicated by the time-independent velocity
width. A linear function with zero slope is fitted to the data with
the uncertainty $\Delta v_{x}^{\prime}=$ 3.47(2) cm s$^{-1}$ and 4.89(6) cm s$^{-1}$, as shown in Fig. 2(a). The error bars in this experiment represent the standard error of the mean calculated from 20 data sets. Using Eq. (\ref{eq:En2})
and the measured uncertainty of the ground state velocity $\Delta v_{0}$
= 2.01(4) cm s$^{-1}$, we can extract the experimental squeezing
amplitudes $r$ = 0.89(6) and 1.24(7) by taking the ratio of $\Delta v_{x}^{\prime}/\Delta v_{0}=\left(\cosh2r\right)^{1/2}$.
The measured squeezing amplitude is smaller than $\ln\left(\omega/\omega^{\prime}\right)$
= 1.21 and 1.75 due to the imperfect ground state cooling, anharmonicity,
and the available bound states \citep{Xin}. When the two input squeezed
states are in-phase ($\theta_{x}-\theta_{y}=0$), the output state
becomes a two-dimensional single-mode squeezed state where the velocity
width oscillates at twice the oscillator frequency, as shown in Fig.
\ref{fig:Fig2}(b). The fitted squeezing amplitude $r$ = 0.89(8) for
$\ln\left(\omega/\omega^{\prime}\right)$ = 1.21 agrees well with
the measurement of the two-mode squeezed state, which corresponds to 7.7 dB of squeezing. The achievable squeezing in our system is limited by the available bound states of the potential. For ln$(\omega/\omega')=1.75$, the available bound states decrease from 15 to 3.

\begin{figure}
\includegraphics[scale=0.3]{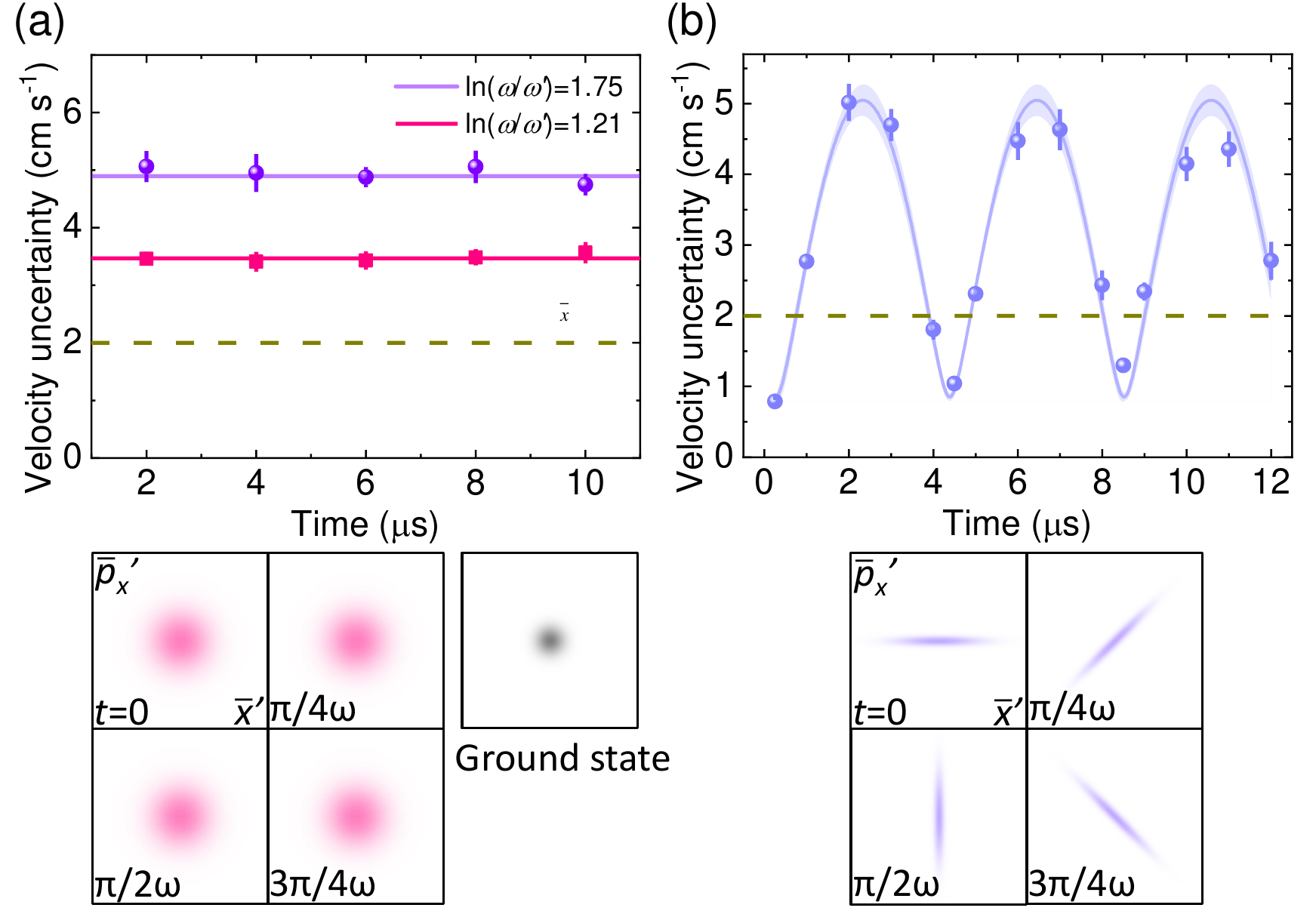}

\caption{\label{fig:Fig2}Characterization of the controlled two-mode squeezing
gate. (a) Measurements of the velocity uncertainty as a function of
free evolution time along the $x^{\prime}$ direction using the Raman
beams $\text{X}_{2}$ and $\text{Y}_{1}$ when the two input squeezed
states are out-of-phase. The red and purple lines are fitted constants
to the data with different squeezing amplitudes $\ln\left(\omega/\omega^{\prime}\right)$
= 1.21 and 1.75, respectively. The dashed line marks the uncertainty
of ground state velocity. The bottom figures are the calculated time-dependent
Wigner functions projected onto $\bar{p}_{x}^{\prime}$--$\bar{x}^{\prime}$
plane with $r$ = 0.89, where $\bar{p}_{x}^{\prime}$ and $\bar{x}^{\prime}$
are dimensionless quadratures. The ground state Wigner function is
plotted for reference. (b) Measurements of the velocity uncertainty
as a function of free evolution time along the $x^{\prime}$ direction
using the Raman beams $\text{X}_{2}$ and $\text{Y}_{1}$ when the
two input squeezed states are in-phase. The blue curve fits the data
using Eq. (\ref{eq:En2}), where the band indicates 68\% confidence
level.}
\end{figure}

Despite our two modes in the two-mode squeezed state remain degenerate, we can still use the Duan-Simon criterion \citep{Dua,Sim} to quantify the state. The two modes are inseparable
when the variance of the difference and sum of the dimensionless quadrature
amplitudes add up to less than one: $\Delta^{2}\left(\bar{x}^{\prime}-\bar{y}^{\prime}\right)+\Delta^{2}\left(\bar{p}_{x}^{\prime}-\bar{p}_{y}^{\prime}\right)<1$,
where $\bar{x}^{\prime}(\bar{y}^{\prime})\equiv x^{\prime}(y^{\prime})/2\Delta x_{0}$,
$\bar{p}_{x}^{\prime}(\bar{p}_{y}^{\prime})\equiv p_{x}^{\prime}(p_{y}^{\prime})/2\Delta p_{0}$,
and $\Delta x_{0}$ is the uncertainty of the ground state wave packet
size. A more stringent criterion that satisfies the EPR criterion
\citep{Rei} requires $\Delta\left(\bar{x}^{\prime}-\bar{y}^{\prime}\right)\Delta\left(\bar{p}_{x}^{\prime}-\bar{p}_{y}^{\prime}\right)<1/4$,
stemming from the Heisenberg uncertainty principle that certifies
steering. Such a condition has been demonstrated with internal spin
states of atoms \citep{Pei,Mat,Kun,Lan,Shi,Col} but has not been fulfilled
in mechanical oscillators \citep{Ock,Kot,Lau}. Although we couldn\textquoteright t
measure $\Delta\left(\bar{x}^{\prime}-\bar{y}^{\prime}\right)$ directly,
we can refer to it from the measurement of $\Delta\left(\bar{p}_{x}^{\prime}-\bar{p}_{y}^{\prime}\right)$
after a quarter period of the free evolution. The uncertainty of the
momentum can be calculated as

\begin{equation}
\begin{split}\frac{\Delta\left(\bar{p}_{x}^{\prime}-\bar{p}_{y}^{\prime}\right)}{\sqrt{2}} & =\Delta \bar{p}_{x}\left(\theta_{x}\right) \\
& =(e^{-2r}\sin^{2}\theta_{x}+e^{2r}\cos^{2}\theta_{x})^{1/2}/2,\\
\frac{\Delta\left(\bar{p}_{x}^{\prime}+\bar{p}_{y}^{\prime}\right)}{\sqrt{2}} & =\Delta \bar{p}_{y}\left(\theta_{x}\right) \\
& =(e^{-2r}\cos^{2}\theta_{x}+e^{2r}\sin^{2}\theta_{x})^{1/2}/2.
\end{split}
\label{eq:En3}
\end{equation}

\begin{figure}
\includegraphics[scale=0.45]{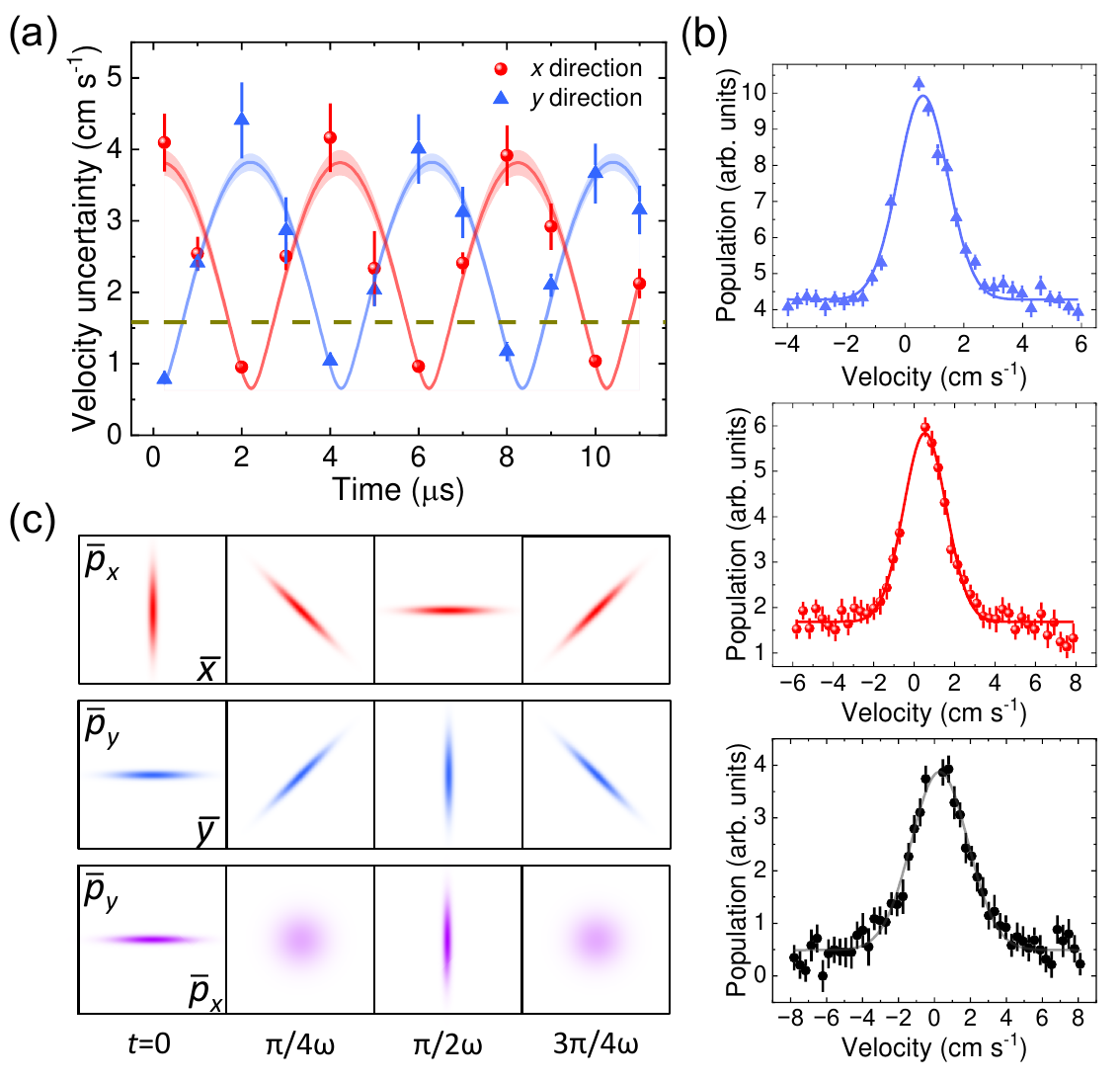}

\caption{\label{fig:Fig3}EPR criterion measurements of the two-mode squeezed
states. (a) Measurements of the velocity uncertainty along the $x$($y$) direction as a function of free evolution time using Raman beams $\text{X}_{1}$($\text{Y}_{1}$) and $\text{X}_{2}$($\text{Y}_{2}$) when the two input squeezed states are out-of-phase. The curves fit the data using Eq. (\ref{eq:En3}), where the band indicates 68\% confidence level. The dashed line marks the uncertainty of ground state velocity. (b) Velocity distribution measurements of the data in Fig. 3(a). Top: The $y$ direction data point of free evolution time at 0.25 $\mu$s. Middle: The $x$ direction data point of free evolution time at 2 $\mu$s. Bottom: The ground state velocity width measurement indicated by a dashed line in Fig. 3(a). The curves are Gaussian functions fitted to the data. (c) Calculated time-dependent Wigner function projected onto $\bar{p}_{x}$--$\bar{x}$, $\bar{p}_{y}$--$\bar{y}$, and $\bar{p}_{y}$--$\bar{p}_{x}$ planes with $r$ = 1.10 , where $\bar{x}$,
$\bar{y}$, $\bar{p}_{x}$, and $\bar{p}_{y}$ are dimensionless quadratures.}
\end{figure}

\begin{figure*}
\includegraphics[scale=0.375]{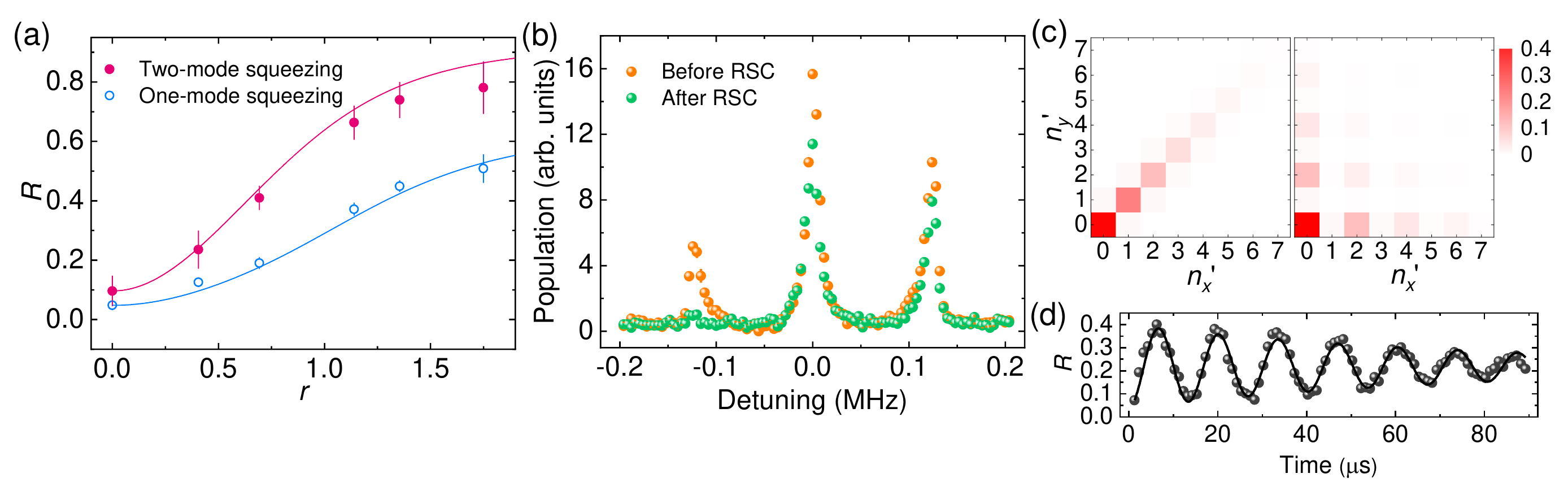}

\caption{\label{fig:Fig4}Analysis of the two-mode squeezed states in the Fock state basis. (a) Measured $R$, the ratio of the first red to first
blue sideband populations, versus squeezing amplitude $r$ with one and two-mode squeezed states. The curves are the theory based on the measured quantities without any free parameters. (b) A typical Lamb-Dicke spectroscopy of the oscillators in the ground state and thermal state (before RSC). The right, center, and left peaks are the first blue sideband, carrier, and first red sideband, respectively. (c) The reconstructed probabilities of occupying a state with $n_{x}^{\prime}$ phonons in the $x^{\prime}$ mode and $n_{y}^{\prime}$ phonons in the $y^{\prime}$ mode from the fitted results of Fig. \ref{fig:Fig2} for the two-mode squeezed state (left) and the two-dimensional single-mode squeezed state (right). (d) Measured $R$ of $\hat{S}^{\dagger}\left(r\right)\hat{U}\left(\omega\tau\right)\hat{S}\left(r\right)$ operation versus the free oscillation time $\tau$ with $r$ = 1.2. The curve is a fit to the data using an exponentially decaying sinusoidal function. The fitted $1/e$ decay constant is 80(9) $\mu$s.}
\end{figure*}

Two pairs of counter-propagating Raman beams along $x$ and $y$ are
used to measure the velocity distribution, as shown in Fig. \ref{fig:Fig1}(a).
The data in Fig. \ref{fig:Fig3}(a) are fitted with Eq. (\ref{eq:En3})
with measured squeezing amplitude $r$ = 0.88(9) for $\ln\left(\omega/\omega^{\prime}\right)$
= 1.21, which is consistent with the measurements in Fig. \ref{fig:Fig2}. The ground state velocity uncertainty is plotted for reference, and the smaller velocity uncertainty compared to the measurements in Fig. 2(b) could be due to the imperfect alignment of the Raman beams for the measurements in Fig. 2(b). The EPR criterion in our system is calculated by taking the first smallest velocity uncertainty data point in the $x$ direction from Fig. 3(a) as $\left.\Delta\left(\bar{p}_{x}^{\prime}-\bar{p}_{y}^{\prime}\right)\right|_{\theta_{x}=\pi/2}$ and that in the $y$ direction as $\left.\Delta\left(\bar{p}_{x}^{\prime}+\bar{p}_{y}^{\prime}\right)\right|_{\theta_{x}=0}$. The results yield $\left.\Delta\left(\bar{p}_{x}^{\prime}-\bar{p}_{y}^{\prime}\right)\right|_{\theta_{x}=\pi/2}\left.\Delta\left(\bar{p}_{x}^{\prime}+\bar{p}_{y}^{\prime}\right)\right|_{\theta_{x}=0}=0.15(3)$, 3 standard deviations below the EPR criterion. The velocity uncertainty measurements of the two points are shown in Fig. 3(b). Figure
\ref{fig:Fig3}(c) plots the calculation of the Wigner function projected
in various coordinate planes versus time for illustration.

In the Fock state representation, a two-mode squeezed state can be
written as $\left|S_{2}\right\rangle =\sech r\Sigma_{n=0}^{\infty}\left(e^{i2\theta_{0}}\tanh r\right)^{n}\left|n,n\right\rangle $
, where $n$ is the number of the Fock state of both modes. Compared
to a single-mode squeezed state where it only contains the even number
of the Fock states, the two-mode squeezed state carries all the Fock
states in both modes. We characterize the state in the Fock state
basis by conducting Lamb Dicke spectroscopy on the state and measuring
the ratio $R$ of the first red to the first blue sideband population
with different squeezing amplitude $r$ as shown in Fig. \ref{fig:Fig4}(a)
and \ref{fig:Fig4}(b). The data is compared with theory (see Supplemental Material),
and the single-mode squeezed state results are also presented in Fig.
\ref{fig:Fig4}(a). The main discrepancy between single-mode and two-mode
states is the variance of their phonon numbers. Despite the same mean
phonon number of both single-mode and two-mode squeezed states, the
variance of phonon number in a single-mode state is two times larger
\citep{Ger}. We further illustrated this by reconstructing the two-dimensional
Fock states probability distributions of the two-mode and two-dimensional
single-mode squeezed states, as shown in Fig. \ref{fig:Fig4}(c), using
the fitted results of the squeezing amplitude and the initial imperfect
ground state from Fig. \ref{fig:Fig2} combined with Eq. (S2) and Eq. (S3) for two-mode squeezed states and Eq. (S6) and Eq. (S7) in Ref. \citep{Xin} for single-mode squeezed states. Figure \ref{fig:Fig4}(d) characterize
the coherence of the single-mode squeezed states by waiting for a free
evolution time $\tau$ in between a unitary operation $\hat{S}^{\dagger}\left(r\right)\hat{U}\left(\omega\tau\right)\hat{S}\left(r\right)$,
where $\hat{S}\left(r\right)$ is the single mode squeezing operator
and $\hat{U}$ is the free evolution operator. The fitted $1/e$ decay
time of the oscillation period of the measured $R$ is 80(9) $\mu$s and is mainly contributed by the inhomogeneous broadening of the oscillation frequency across different lattice sites,
matching well with the measured linewidth in the Lamb-Dicke spectroscopy
shown in Fig. \ref{fig:Fig4}(b).

Our demonstration of the interaction-free generation of a two-mode
squeezed state can be applied in quantum information processing combined
with other operations. By modulating one of the outputs with a displacement
operator and recombining the two outputs, a quantum dense coding scheme
\citep{Bra2} can be realised in mechanical oscillators. An unknown
state from a third mode can be teleported by combining it with one
of the modes on a beam splitter \citep{Bra3}. The measurement result
of the output state is then used to perform a displacement operation
to obtain the unknown state on the other mode. Non-Gaussian operations
such as photon subtractions have many applications in the continuous-variable
approach and the generation of non-classical states. Taking advantage
of discrete variables inherited from atomic oscillators, such as spins,
single-phonon subtraction can be accomplished through spin-oscillator
coupling \citep{Um} as a phonon counting measurement. Preparation
of cubic phase gate by displacement operation and phonon counting
on a two-mode squeezed state has also been proposed \citep{Sho}.
Finally, implementation of spin-dependent optical lattices along different modes would allow to separate two distinct modes spatially, enabling a test of non-locality of massive particles using external degrees of freedom \citep{Bru}.

\begin{acknowledgments}
This work was financially supported by Singapore National Research
Foundation under grant number NRF2021-QEP2-03-P01 and QEP-P4, and Singapore
Ministry of Education under grant number MOE-T2EP50121-0021. Shau-Yu Lan acknowledges the support of the Yushan Fellow Program by the Ministry of Education (MOE), Taiwan, and 2030 Cross-Generation Young Scholars Program by the National Science and Technology Council (NSTC), Taiwan, under grant number 112-2628-M-002-013-.
\end{acknowledgments}

\bibliographystyle{apsrev4-2}
\nocite{*}
\bibliography{TMS}

\maketitle
\section*{Supplemental Material}
\section{Preparation of the quantum registers in a two-dimensional optical
lattice}
When cold atoms are loaded into the optical lattice, we perform polarization
gradient cooling (PGC) for 2 ms and then switch off the Y lattice
beam for a quarter period of radial motion to compress the atoms along
$y$ in the X lattice beam. After the atoms are compressed to the
bottom of the X trap, we switch on the Y lattice beam. The procedure
is repeated for the X lattice beam to compress the atoms into a small
number of two-dimensional lattice tubes, followed by another PGC.

We perform resolved Raman sideband cooling (RSC) on the $\left|F=2,m_{F}=0\right\rangle $
state to prepare atoms in the two-dimensional vibrational ground state [16], where $F$ denotes the hyperfine ground state of $^{85}$Rb
and $m_{F}$ is the Zeeman state. Two pairs of Raman lasers, as shown
in Fig. 1(a), are 200 GHz red-detuned from the $\left|F=2\right\rangle $
to $\left|F^{\prime}=2\right\rangle $ transition in the D1 line with
2 mW of the Raman beams $\text{X}_{1}$($\text{Y}_{1}$) and 100 mW
of the Raman beams $\text{X}_{2}$($\text{Y}_{2}$), where $F^{\prime}$
is the hyperfine excited state. The imbalance of the Raman beams is due to technical constraint. The atoms are cooled in both dimensions
by performing alternating RSC cycles in the $x$ and $y$ directions.
Each cooling cycle consists of 100 $\mu$s of the Raman pulses and
50 $\mu$s of the depump and optical pump pulses. The overall duration
for the best cooling performance is 320 cooling cycles. After RSC,
$\sim$75\% of the atoms remain in the trap and the mean vibrational
quantum number $\bar{n}_{0}$ = 0.06(4) is characterized using Lamb-Dicke
spectroscopy along the $x^{\prime}$ direction. Figure 5 shows the spectroscopy along different directions.

We prepare the individual squeezed states through sudden jumps of
the oscillator frequency from $\omega$ to $\omega^{\prime}$, which
transforms the vibrational ground state into a squeezed state in the
new eigenstates basis. After a free oscillation time $\pi/\left(2\omega^{\prime}\right)$,
the squeezed states rotate at an angle of $\pi/2$ and the oscillator
frequency is immediately switched back to $\omega$ to further squeeze
the atomic wave function. The oscillator frequency is controlled by
modulating the lattice beams amplitude using acousto-optic modulators
(AOMs).

\begin{figure}
\includegraphics[width=0.9\columnwidth]{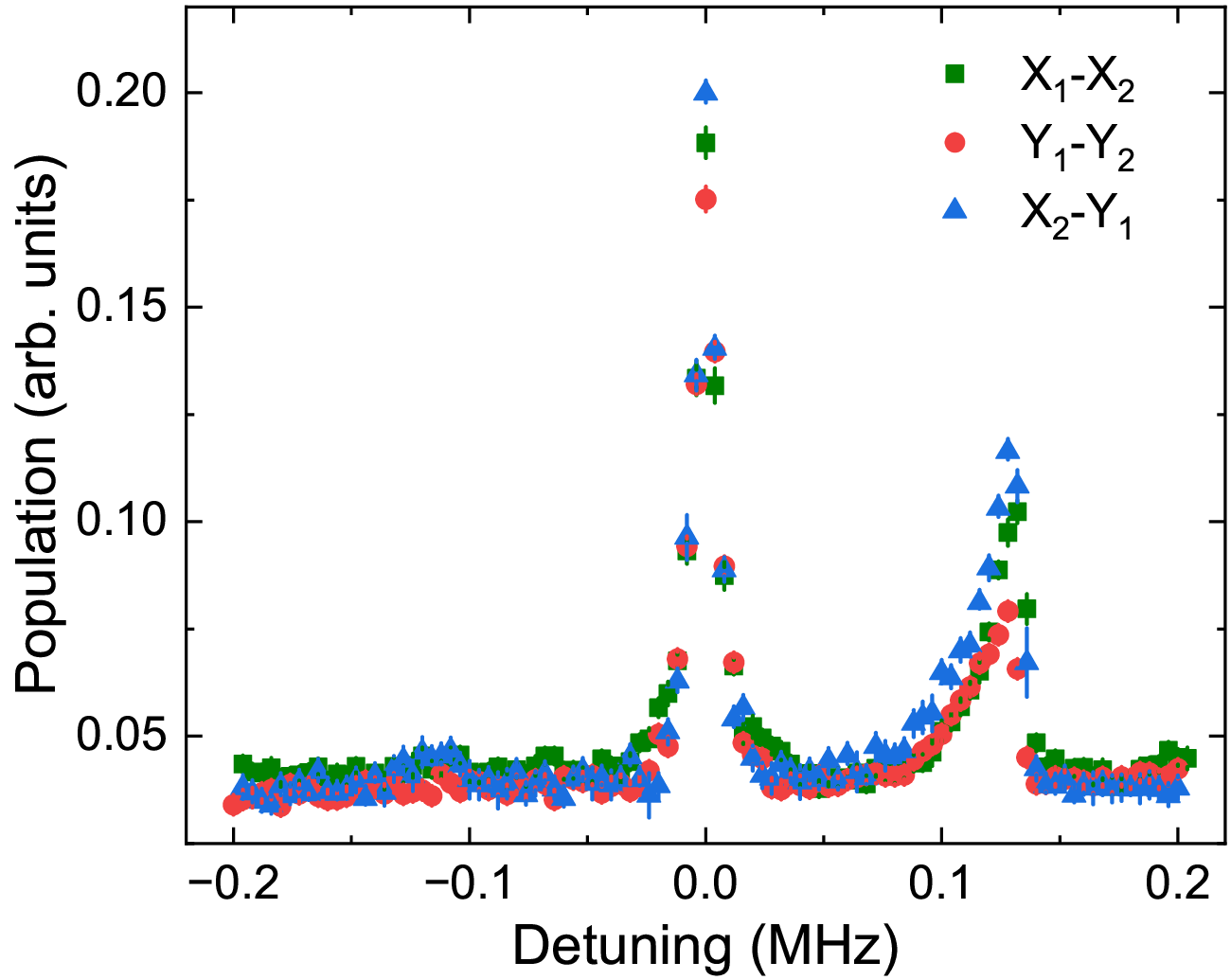}

\caption{\label{fig:FigS1}Lamb‐Dicke spectroscopy after RSC along different directions. The measurements are performed using Raman beams X$_{1(2)}$ and Y$_{1(2)}$.}
\end{figure}

\section{Velocity width measurement}
The velocity width of the atomic wave packet is inferred from the
measurement of the velocity-sensitive two-photon Raman process after a sudden release of the atoms from the lattice. The
Raman beams $\text{X}_{1}$($\text{Y}_{1}$) and $\text{X}_{2}$($\text{Y}_{2}$)
are used to measure the width along $x$($y$), while the Raman beams
$\text{X}_{2}$ and $\text{Y}_{1}$ are used to measure the width
along $x^{\prime}$. A pair of the Raman beams, shown in Fig. 1(a),
couple $\left|F=2\right\rangle $ and $\left|F=3\right\rangle $ to
form a two-photon Raman transition. They excite atoms from $\left|F=2\right\rangle $
to $\left|F=3\right\rangle $ when the relative frequency detuning
of the two lasers, which is controlled using a double-passed 1.5 GHz
AOM, satisfies the hyperfine splitting $\omega_{\text{HF}}$. At a
low optical power limit, the velocity width of the atoms that are
excited to the $\left|F=3\right\rangle $ state is determined by the
duration of the square pulse \cite{Kas}. Therefore, by scanning
the relative frequency detuning the lasers, the velocity distribution
of the atoms can be mapped out. The AOM for controlling the lattice optical power exhibits a switching off time of approximately 200 ns, which is significantly shorter than our minimum oscillation period of 8 $\mu$s to ensure the motion of atoms in the trap do not affect the results of the measurements. The velocity uncertainty is then determined
by fitting a Gaussian function to the data. We use a pulse width of
0.1 ms though out the measurements. Figure 6 shows two velocity width measurements of the data points in Fig. 2(b) and the ground state velocity width measurements.

\begin{figure}
\includegraphics[width=0.8\columnwidth]{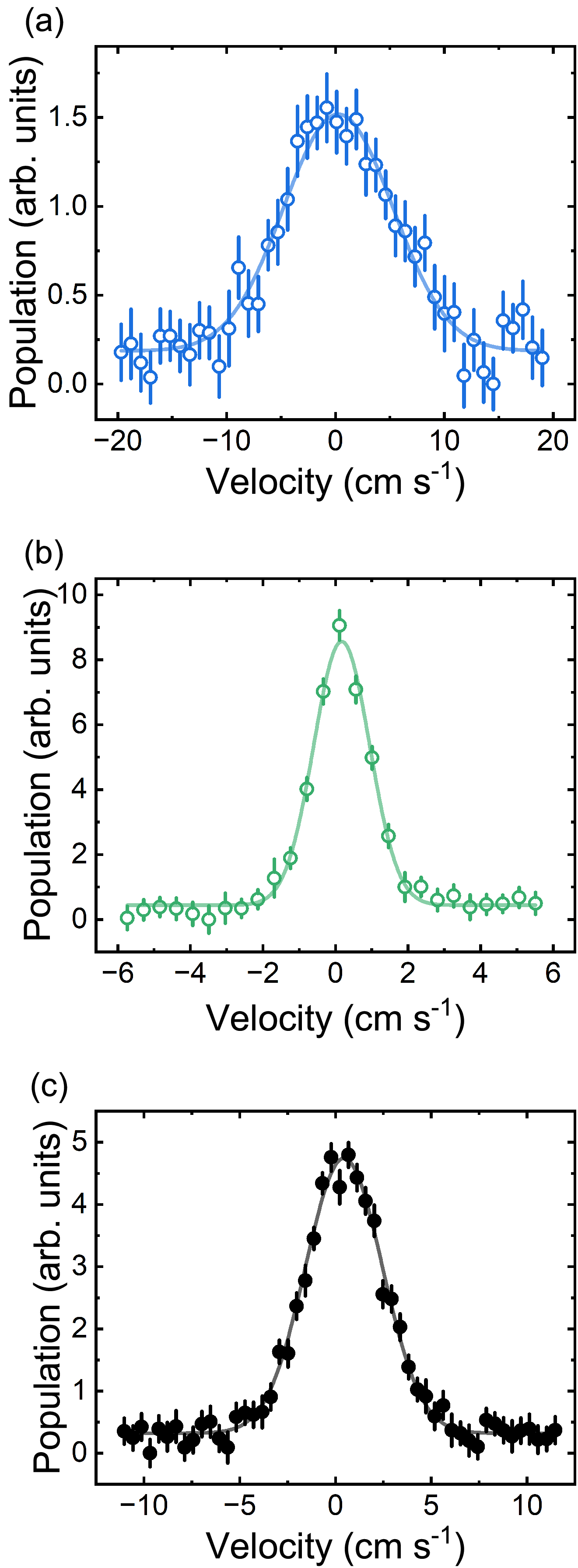}

\caption{\label{fig:FigS2}Velocity distribution measurements of the data in Fig. 2(b). (a) and (b) correspond to the data points of free evolution time at 0.25 and 2 $\mu$s, respectively. (c) is the ground state velocity width measurements indicated as a dashed line in Fig. 2(b). The curves are Gaussian functions fitted to the data.}
\end{figure}

\section{Analysis of the two-mode squeezed state in the Fock state basis}
We characterize the two-mode squeezed state by analyzing their Fock
state population distribution. This is carried out by taking the ratio
$R$ of the measured first red sideband peak population to the first
blue sideband peak population in the Lamb-Dicke spectrum, as shown in Fig. 4c. In the spectrum, the Raman beams $\text{X}_{1}$ and $\text{Y}_{2}$ excite the atoms from$\left|F=2,n_{x}^{\prime},n_{y}^{\prime}\right\rangle $ to$\left|F=3,n_{x}^{\prime}-1,n_{y}^{\prime}\right\rangle $ (the first red sideband $P_{-}$) when the relative frequency detuning of the two lasers satisfies $\omega_{\text{HF}}-\omega$. When the two lasers are frequency detuned by $\omega_{\text{HF}}+\omega$, the atoms are populated into$\left|F=3,n_{x}^{\prime}+1,n_{y}^{\prime}\right\rangle $
(the first blue sideband $P_{+}$). The population of each sideband after a Raman pulse duration $t$ is proportional to the sum of the probability in each Fock state after Rabi flopping as
\begin{equation}
\begin{split}P_{+}\left(t\right) & \propto\sum_{n_{x}^{\prime},n_{y}^{\prime}=0}^{\infty}P_{n_{x}^{\prime},n_{y}^{\prime}}\frac{1-e^{-\gamma t}\cos\left(\sqrt{n_{x}^{\prime}+1}\Omega_{0,1}t\right)}{2}\\
P_{-}\left(t\right) & \propto\sum_{n_{x}^{\prime},n_{y}^{\prime}=0}^{\infty}P_{n_{x}^{\prime},n_{y}^{\prime}}\frac{1-e^{-\gamma t}\cos\left(\sqrt{n_{x}^{\prime}}\Omega_{0,1}t\right)}{2},
\end{split}
\label{eq:En4}
\end{equation}
where $P_{n_{x}^{\prime},n_{y}^{\prime}}$ is the probability in the state $\left|F=2,n_{x}^{\prime},n_{y}^{\prime}\right\rangle $, $\Omega_{0,1}=\eta\Omega$ is the two-photon Rabi frequency of $\left|F=2,n_{x}^{\prime}=0,n_{y}^{\prime}\right\rangle $
to $\left|F=3,n_{x}^{\prime}=1,n_{y}^{\prime}\right\rangle $ transition, $\eta=0.13$ is the Lamb Dicke parameter, $\Omega$ is the two-photon Rabi frequency of the carrier transition, and $\gamma$ is the decay rate of the Rabi flopping. We observe that the primary factor causing decoherence in the Fock states is the inhomogeneous broadening of the vibrational frequency across various lattice sites. As a result, we attribute a uniform decay rate to all Fock states. We calculate the expected population $P_{\pm}\left(t\right)$ and compare it with our measurements. In the calculation, $\Omega_{0,1}$= 2$\pi$ $\times$ 1.5 kHz and $\gamma$ = 10.36 kHz are determined from the measured Rabi flopping. The pulse duration $t=0.17$ ms is deliberately selected to be significantly shorter than the scattering time of approximately 1 ms caused by the Raman beams for single photons, while still being longer than a single Rabi oscillation cycle. The sum replaces the infinite sum up to the term $n_{x}^{\prime}$ = 25, $n_{y}^{\prime}$ = 25 in our calculation due to computation limitation, and these numbers are more than enough for our experiment. The probability $P_{n_{x}^{\prime},n_{y}^{\prime}}$ takes into account the imperfect RSC by weighing the Boltzmann distribution as
\begin{equation}
\begin{split}P_{n_{x}^{\prime},n_{y}^{\prime}} & =\sum_{l_{x}^{\prime},l_{y}^{\prime}=0}^{\infty}\frac{1}{\left(1+\bar{n}_{0}\right)^{2}}\left(\frac{\bar{n}_{0}}{1+\bar{n}_{0}}\right)^{l_{x}^{\prime}+l_{y}^{\prime}}\\
 & \qquad\times\left|\left\langle n_{x}^{\prime},n_{y}^{\prime}\left|\hat{S}_{2}\right|l_{x}^{\prime},l_{y}^{\prime}\right\rangle \right|^{2},
\end{split}
\label{eq:En5}
\end{equation}
where the squeezing operator $\hat{S}_{2}\left(r\right)$ can be calculated as \cite{Kim}
\begin{widetext}
\begin{equation}
\begin{split}\left|\left\langle n_{x}^{\prime},n_{y}^{\prime}\left|\hat{S}_{2}\right|l_{x}^{\prime},l_{y}^{\prime}\right\rangle \right|^{2} & =\delta_{l_{y}^{\prime}-n_{y}^{\prime}}^{l_{x}^{\prime}-n_{x}^{\prime}}\frac{\tanh^{2\left(l_{x}^{\prime}-n_{x}^{\prime}\right)}r}{\cosh^{2\left(n_{x}^{\prime}+n_{y}^{\prime}+1\right)}r}\left|\sum_{g=0}^{\min\left(n_{x}^{\prime},n_{y}^{\prime}\right)}\left(-\sinh^{2}r\right)^{g}\sqrt{\binom{n_{y}^{\prime}}{g}\binom{l_{x}^{\prime}}{l_{x}^{\prime}-n_{x}^{\prime}+g}\binom{l_{y}^{\prime}}{l_{y}^{\prime}-n_{y}^{\prime}+g}}\right|^{2}\end{split}
\label{eq:En6}
\end{equation}
\end{widetext}
$\delta$ is the Kronecker delta, and $\binom{*}{*}$ is the binomial
coefficient. Here, the infinite sum is replaced by the sum up to the
term $l_{x}^{\prime}+l_{y}^{\prime}\leq50$ in our calculation.

\end{document}